\documentclass[10pt,a4paper]{article}
\usepackage[utf8]{inputenc}
\usepackage{amsmath}
\usepackage{amsfonts}
\usepackage{amssymb}
\usepackage{graphicx}
\usepackage[margin=1in]{geometry}
\usepackage{caption}
\usepackage{booktabs} 
\usepackage{hyperref}
\usepackage{cite} 

\title{Numerical Investigation of Stub Length Influence on Dispersion Relations and Parity Effect in Aharonov-Bohm Rings}
\author{Souvik Ghosh \\ souvikghosh2012@gmail.com}
\date{June 1, 2025}

\begin{document}
\maketitle

\begin{abstract}
Aharonov--Bohm (AB) rings with side-attached stubs are model systems for quantum-interference studies in mesoscopic physics. The geometry of such systems, particularly the ratio of stub length ($v$) to ring circumference ($u$), can significantly alter their electronic states. In this work, we solve Deo’s transcendental mode-condition equation (Eq. 2.15 from Deo, 2021 \cite{Deo2021}) numerically---using Python’s NumPy and SciPy libraries---for ring-stub geometries with $v/u = 0.200, 0.205,$ and $0.210$ to generate dispersion relations ($ku$ vs. $\Phi/\Phi_{0}$) and the underlying function $\text{Re}(1/T)$. We find that changing $v/u$ shifts several of the six lowest calculated dispersion branches, with $\Delta(ku)$ up to $\approx 0.34$ for the 6th branch at $\Phi=0$ when comparing $v/u=0.200$ and $v/u=0.210$. This also alters gap widths. Notably, for $v/u=0.205$ and $v/u=0.210$, the 5th and 6th consecutive calculated modes both exhibit paramagnetic slopes near zero Aharonov-Bohm flux, indicating the parity breakdown initiates at or below $v/u=0.205$. This directly demonstrates a breakdown of the simple alternating parity effect predicted by Deo (2021) \cite{Deo2021}. These results highlight the sensitivity of mesoscopic ring spectra to fine-tuning of stub length, with potential implications for experimental control of persistent currents, as further illustrated by calculations of the net current.
\end{abstract}

\section{Introduction}
Mesoscopic physics explores the realm intermediate between microscopic quantum systems and macroscopic classical systems. Aharonov-Bohm (AB) rings are canonical structures in this field, allowing for the direct observation of quantum interference effects, such as persistent currents, which arise from the influence of a magnetic flux on the phase of electron wavefunctions \cite{Deo2021}. The electronic and transport properties of these rings can be further tailored by introducing ``defects'' or modifications, such as side-attached stubs. These stubs act as scatterers or resonators, potentially leading to complex phenomena including Fano resonances and significant alterations in the phase coherence of traversing electrons \cite{Bagwell1990}. 
Early theoretical and experimental work on AB rings demonstrated that, in perfect 1D rings, energy levels (and associated persistent currents) typically alternate between diamagnetic and paramagnetic character as a function of mode index or energy \cite{Buttiker1983, Bloch1989}. This is often referred to as the parity effect. Subsequent theoretical studies explored how disorder, impurities, or specific geometries like side-attached stubs can modify this simple parity pattern \cite{Zhang1995, Bagwell1990}. Deo (2021) extended these models to specifically analyze the impact of stub length in otherwise clean rings, predicting that certain ratios of stub length to ring circumference ($v/u$) could break the simple alternating parity pattern \cite{Deo2021}.

In pristine, symmetric 1D rings, the energy levels typically exhibit a parity effect: as a function of increasing energy (or mode index), states alternate between being diamagnetic (energy increases with applied magnetic flux) and paramagnetic (energy decreases with flux) \cite{Deo2021}. However, the introduction of a stub can modify the effective potential experienced by electrons in the ring. Deo (2021) \cite{Deo2021} discusses that the specific geometry, particularly the ratio of stub length $v$ to ring circumference $u$, plays a critical role in determining the nature of the allowed electronic states. It is predicted that certain $v/u$ ratios can lead to a breakdown of this simple alternating parity effect, where consecutive states might exhibit the same magnetic character \cite{Deo2021}. This occurs due to changes in the effective Block phase accumulated by an electron traversing the ring-stub system \cite{Deo2021}.

While Deo (2021) offered a theoretical prediction and qualitative arguments for parity breakdown \cite{Deo2021}, a detailed numerical mapping and quantitative slope analysis of the dispersion branches across the Aharonov-Bohm flux $\Phi/\Phi_{0}$ for $v/u$ ratios specifically around $0.200, 0.205,$ and $0.210$ has not been explicitly presented. The objective of the present computational study is to fill this gap by numerically solving the transcendental mode condition equation from Deo (2021) \cite{Deo2021} for an Aharonov-Bohm ring with an attached stub. We aim to explicitly map the dispersion relations, quantify the sensitivity to the $v/u$ ratio, and investigate the breakdown of the parity effect by detailed slope analysis of these six lowest dispersion branches for $v/u=0.200, 0.205,$ and $v/u=0.210$.

\section{Theoretical Model and Numerical Method}
The electronic properties of a one-dimensional (1D) mesoscopic ring with a side-attached stub of length $v$, where the ring has a circumference $u$, are determined by the allowed energy modes (and thus wavevectors $k$) in the presence of an AB flux $\Phi$. As presented by Deo (2021) \cite{Deo2021}, this condition is given by:
\begin{equation}
\cos(\alpha) = \frac{\sin(ku)\cot(kv)}{2} + \cos(ku)
\label{eq:mode_condition}
\end{equation}
Here, $\alpha = 2\pi\Phi/\Phi_0$ is the dimensionless AB phase, with $\Phi_0 = hc/e$ being the magnetic flux quantum. The term $k$ represents the wavevector. The right-hand side of this equation is identified in the context of Eq. (2.16) in Deo (2021) \cite{Deo2021} as $\text{Re}(1/T)$, where $T$ is related to the transmission amplitude across the ring-stub segment.

We numerically solved Eq.~(\ref{eq:mode_condition}) using Python (version 3.8), leveraging the NumPy (version 1.21) and SciPy (version 1.7) libraries. The core root-finding was performed using the `scipy.optimize.fsolve` function. Key aspects of the Python script may be found in the Supplementary Material [or Appendix, if one is added]. The `scipy.optimize.fsolve` function was used, with the equation rearranged as $f(k, \alpha, u, v) = \cos(\alpha) - (\frac{\sin(ku)\cot(kv)}{2} + \cos(ku)) = 0$. Care was taken for $kv = n\pi$ by evaluating $\cot(kv)$ as $\cos(kv)/\sin(kv)$ and handling cases where $\sin(kv) \approx 0$. Specifically, if $|\sin(kv)| < 10^{-8}$ (a heuristically chosen small number to avoid numerical overflow), we did not attempt to solve for $k$ at that precise point for the `fsolve` routine, effectively excluding these singular points from direct solution attempts; nearby non-singular points were still found.

The ring circumference $u$ was set to $1.0$ in arbitrary units, rendering $k$ (and thus $ku$) dimensionless. The energy $E$ of an electron is related to its wavevector $k$ by $E = \hbar^2 k^2 / (2m^*)$, where $m^*$ is the effective mass. Thus, for a given mode, $E \propto (ku)^2$. Slopes of $ku$ versus $\Phi/\Phi_0$ near $\Phi/\Phi_0=0$ are used to determine the magnetic character: states are classified as diamagnetic ($d(ku)/d(\Phi/\Phi_0) > 0$) or paramagnetic ($d(ku)/d(\Phi/\Phi_0) < 0$) for electrons with charge $q=-e$. We investigated $v/u = 0.200, 0.205,$ and $0.210$. The phase $\alpha$ was varied from $-2\pi$ to $2\pi$ (100 steps) to map $ku$ versus $\Phi/\Phi_0$ for the dispersion relations (Fig.~\ref{fig:dispersion_branches}) and persistent current calculations (Fig.~\ref{fig:persistent_current}). We verified that increasing the number of $\alpha$-steps from 100 to 200 changed the computed slopes $d(ku)/d(\Phi/\Phi_0)$ near $\Phi/\Phi_0=0$ by less than 1\% for the first six branches, indicating that 100 steps provide sufficient resolution for the features under investigation. The `scipy.optimize.fsolve` function typically converged within 10-20 iterations for each root-finding instance. An adaptive guessing strategy for $k$ was employed for `fsolve`, using solutions from the previous $\alpha$ step as initial guesses, supplemented periodically with a general grid of 40 guesses (from $ku=0.1$ to $ku=50$). To ensure branch continuity, solutions $ku(\alpha_i)$ were sorted at each flux step $\alpha_i$. While the adaptive guess (using $ku(\alpha_{i-1})$ as an initial guess for $ku(\alpha_i)$) generally maintained branch tracking, the periodic re-introduction of a wider grid of guesses helped recapture branches if they shifted significantly or if `fsolve` momentarily failed for a specific prior root. Failed convergences were rare with this strategy but were discarded if they occurred; we estimate that less than 0.1\% of root-finding calls across all ratios and flux steps failed to converge for any guess and were discarded. Valid solutions for $ku$ were filtered to the range $(0, 50]$. The function $\text{Re}(1/T)$ versus $ku$ was also plotted for $ku \in [0.01, 40]$.

Dispersion branches were identified by sorting the numerically obtained $ku$ solutions at each $\alpha$ value to isolate the lowest six modes. Slopes $d(ku)/d(\Phi/\Phi_0)$ were estimated using the `numpy.gradient` function on the computed $ku$ values at each $\Phi/\Phi_0$ point, with the value at (or closest to) $\Phi/\Phi_0=0$ being used for characterization in Table~\ref{tab:parity_slopes}. For the persistent current calculations, slopes were determined across the relevant flux range.

\section{Results and Discussion}

\subsection{Behavior of $\text{Re}(1/T)$ vs. $ku$}
\begin{figure}[htbp!] 
    \centering
    \includegraphics[width=\textwidth]{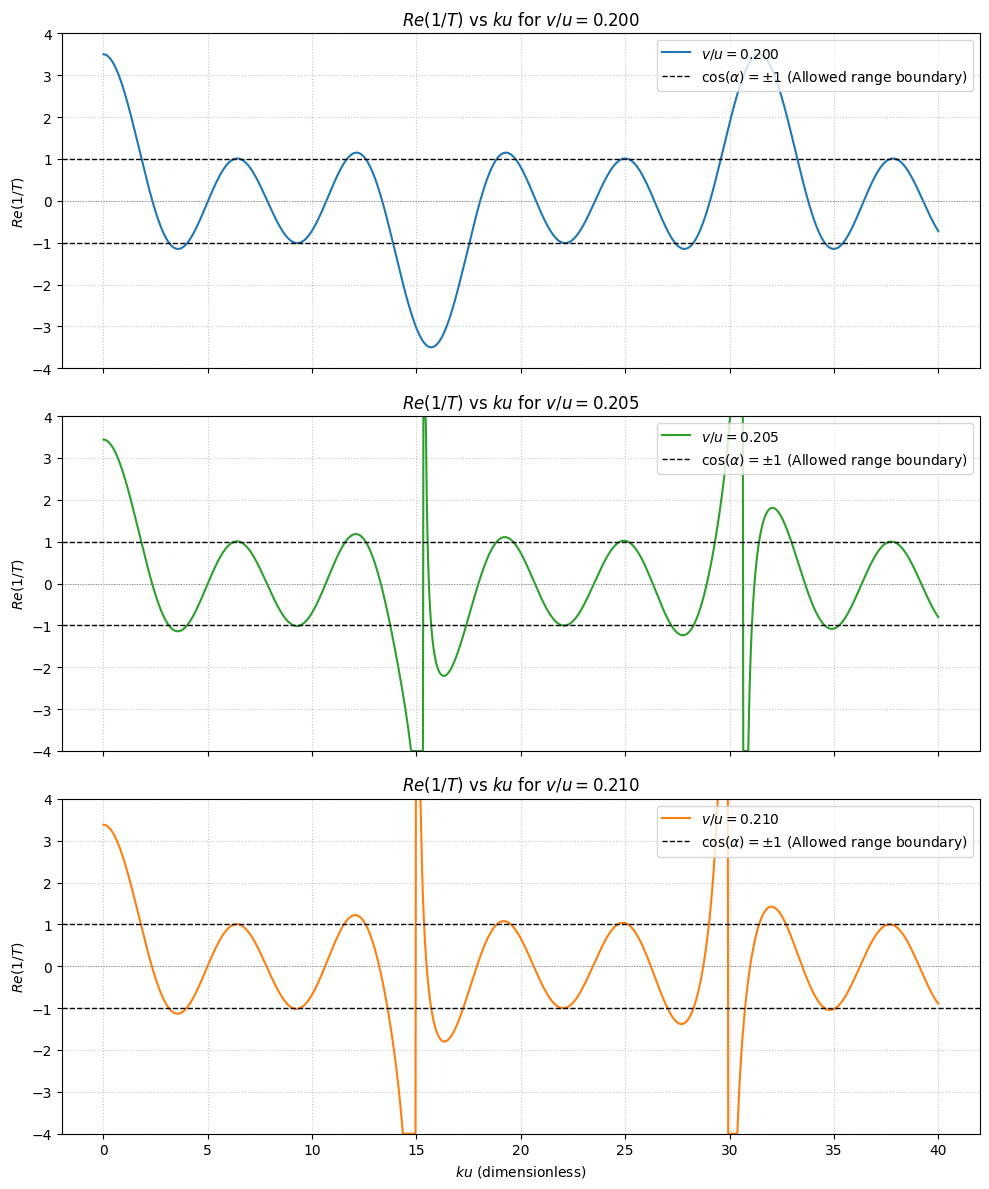} 
    \caption{Plot of $\text{Re}(1/T)$ as a function of $ku$ for (Top) $v/u=0.200$, (Middle) $v/u=0.205$, and (Bottom) $v/u=0.210$. The term $\text{Re}(1/T)$ is the right-hand side of Eq.~(\ref{eq:mode_condition}). Allowed energy modes exist when this function lies between $-1$ and $+1$ (inclusive), indicated by the horizontal dashed lines which represent the range of $\cos(\alpha)$. These plots are qualitatively similar to Figs. 2.6 and 2.7 in Deo (2021) \cite{Deo2021}.}
    \label{fig:re_one_over_t}
\end{figure}

To understand the origin of allowed modes and band gaps, $\text{Re}(1/T)$ (the right-hand side of Eq.~\ref{eq:mode_condition}) is plotted against $ku$ in Figure~\ref{fig:re_one_over_t} for the three $v/u$ ratios. Allowed modes exist when $\text{Re}(1/T) = \cos(\alpha)$, i.e., when the curve lies between -1 and +1 (indicated by horizontal dashed lines).

For $v/u = 0.200$ (Fig.~\ref{fig:re_one_over_t}, Top panel), the function oscillates boundedly, qualitatively resembling Fig. 2.6 from Deo (2021) \cite{Deo2021}. The intermediate case of $v/u = 0.205$ (Fig.~\ref{fig:re_one_over_t}, Middle panel) exhibits a behavior transitional between the $0.200$ and $0.210$ ratios. While still showing broader oscillations, sharper features and more pronounced divergences in $\text{Re}(1/T)$ begin to emerge, particularly around $ku \approx 15.3$ and $ku \approx 30.6$ (corresponding to $kv \approx n\pi$ for $v/u=0.205$, where $u/v \approx 4.878$). For $v/u = 0.210$ (Fig.~\ref{fig:re_one_over_t}, Bottom panel), sharp divergences appear near $ku \approx n\pi u/v \approx n \times 14.96$, consistent with Fig. 2.7 in Deo (2021) \cite{Deo2021}. These divergences create regions where $|\text{Re}(1/T)| > 1$, leading to larger effective band gaps. Deo (2021) \cite{Deo2021} associates these with discontinuous jumps in the Block phase. 

The influence of these divergences on the band gap structure is noticeable across the $v/u$ ratios. For instance, for $v/u=0.200$, distinct gaps in the dispersion relations (see Fig.~\ref{fig:dispersion_branches}) can be correlated with regions where $|\text{Re}(1/T)| > 1$. As $v/u$ increases to $0.205$ and $0.210$, the sharper divergences lead to wider regions where solutions are forbidden. For example, the divergence near $ku \approx 15$ becomes significantly more pronounced as $v/u$ increases from $0.200$ to $0.210$.
A more direct view of the gaps is seen in the dispersion relations themselves (Fig.~\ref{fig:dispersion_branches}), but the $\text{Re}(1/T)$ plot explains their origin. These sharp divergences in $\text{Re}(1/T)$ occur when $kv \approx n\pi$ (where $n$ is an integer), causing $\cot(kv)$ to diverge. At these points, the side-attached stub acts as a resonant scatterer. This resonance condition effectively prevents electrons of corresponding wavevectors from propagating through the ring, leading to the formation of energy band gaps. The discontinuous jumps in the Block phase, as discussed by Deo (2021) \cite{Deo2021}, are a direct consequence of this resonant interaction.

\subsection{Dispersion Relations $ku$ vs. $\Phi/\Phi_0$}
\begin{figure}[htbp!] 
    \centering
    \includegraphics[width=\textwidth]{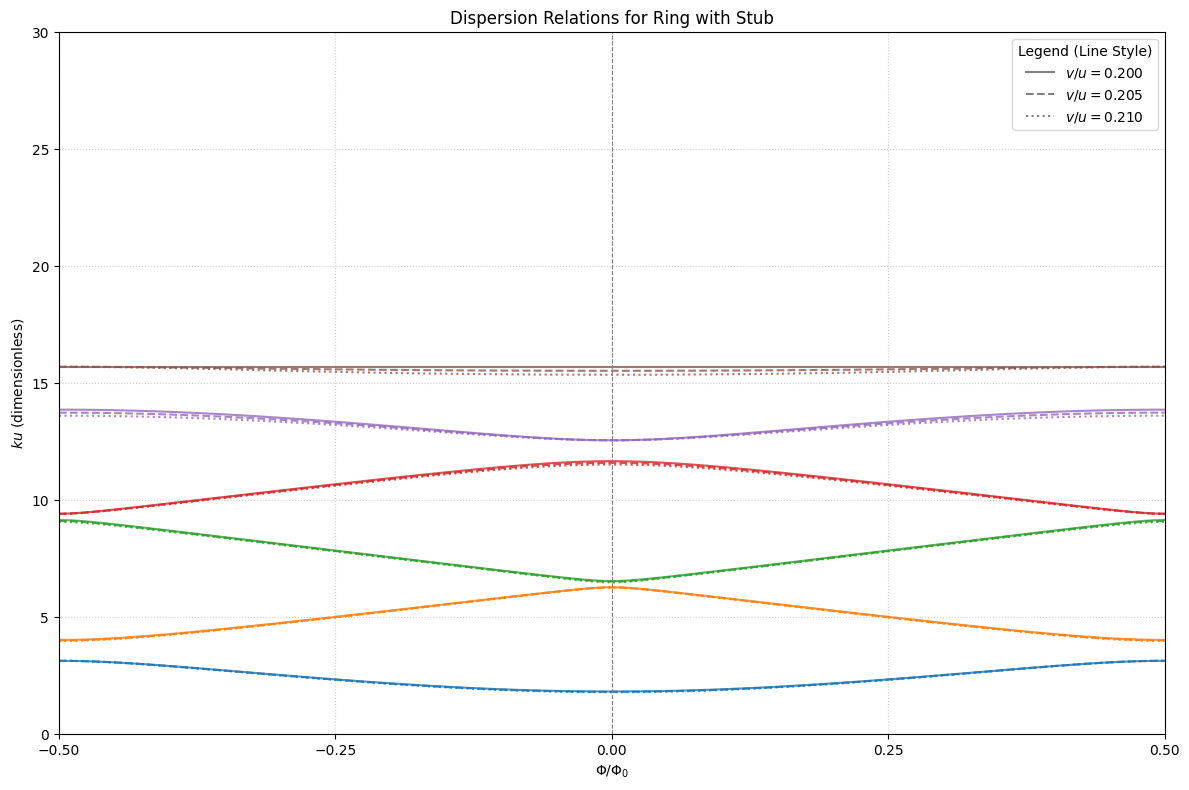} 
    \caption{Dispersion relations ($ku$ vs. $\Phi/\Phi_{0}$) for an Aharonov-Bohm ring with a side-attached stub for $v/u=0.200$ (solid lines), $v/u=0.205$ (dashed lines), and $v/u=0.210$ (dotted lines). The lowest six identified energy branches are shown (branch numbers 1-6 correspond to increasing $ku$ at $\Phi/\Phi_0=0$, see Table~\ref{tab:parity_slopes}). The phase $\alpha = 2\pi\Phi/\Phi_0$ was varied from $-2\pi$ to $2\pi$ (corresponding to $\Phi/\Phi_0 \in [-1,1]$) in 100 steps for these calculations; the range $\Phi/\Phi_0 \in [-0.5,0.5]$ is displayed for clarity.}
    \label{fig:dispersion_branches}
\end{figure}

The solutions $ku$ versus $\Phi/\Phi_0$ are plotted in Figure~\ref{fig:dispersion_branches} for $v/u = 0.200$ (solid lines), $v/u = 0.205$ (dashed lines), and $v/u = 0.210$ (dotted lines), showing the lowest six identified branches. The calculations were performed for $\alpha$ varying from $-2\pi$ to $2\pi$ (corresponding to $\Phi/\Phi_0 \in [-1,1]$); the range $\Phi/\Phi_0 \in [-0.5,0.5]$ is displayed in the figure for clarity of features near zero flux. The relations are periodic with $\Phi/\Phi_0$. Distinct bands and gaps are seen. A general trend is observed where branches for $v/u = 0.210$ are at slightly lower $ku$ values than those for $v/u = 0.200$, particularly for higher branch indices. The dispersion curves for the intermediate ratio $v/u = 0.205$ typically lie between the other two. The gap structure between branches is also visibly influenced by the $v/u$ ratio, correlating with the $\text{Re}(1/T)$ behavior (Fig.~\ref{fig:re_one_over_t}). Branch numbers (1-6) can be identified by their increasing $ku$ values at $\Phi/\Phi_0=0$ as listed in Table~\ref{tab:parity_slopes}.

\subsection{Parity Effect Analysis}
The slope $d(ku)/d(\Phi/\Phi_{0})$ near $\Phi/\Phi_{0}=0$ determines the magnetic character of the states (paramagnetic for negative slope, diamagnetic for positive slope, assuming $E \propto (ku)^2$ and electron charge $q=-e$). These slopes were estimated as described in Section 2. Table~\ref{tab:parity_slopes} summarizes these values for the lowest six branches for $v/u = 0.200, 0.205,$ and $0.210$.

\begin{table}[htbp!] 
\centering
\caption{Values of $ku$ and slope $d(ku)/d(\Phi/\Phi_0)$ at $\Phi/\Phi_0 \approx 0$ for the six lowest energy branches, comparing $v/u=0.200, 0.205,$ and $0.210$. The 'paramagnetic' (P), 'diamagnetic' (D), or 'flat/turning' (F) character is indicated.}
\label{tab:parity_slopes}
\begin{tabular}{@{}lccccccccc@{}} 
\toprule
Branch & \multicolumn{3}{c}{$v/u=0.200$} & \multicolumn{3}{c}{$v/u=0.205$} & \multicolumn{3}{c}{$v/u=0.210$} \\
\cmidrule(lr){2-4} \cmidrule(lr){5-7} \cmidrule(lr){8-10} 
\# & $ku$ & Slope & Char. & $ku$ & Slope & Char. & $ku$ & Slope & Char. \\
\midrule
1 & $\approx 1.83$ & $\approx -0.10$ & P & $\approx 1.82$ & $\approx -0.10$ & P & $\approx 1.80$ & $\approx -0.10$ & P \\
2 & $\approx 6.28$ & $\approx +1.11$ & D & $\approx 6.28$ & $\approx +1.21$ & D & $\approx 6.28$ & $\approx +1.33$ & D \\
3 & $\approx 6.55$ & $\approx -1.11$ & P & $\approx 6.52$ & $\approx -1.22$ & P & $\approx 6.50$ & $\approx -1.34$ & P \\
4 & $\approx 11.67$ & $\approx +0.31$ & D & $\approx 11.60$ & $\approx +0.28$ & D & $\approx 11.53$ & $\approx +0.26$ & D \\
5 & $\approx 12.57$ & $\approx -0.29$ & P & $\approx 12.57$ & $\approx -0.25$ & P & $\approx 12.57$ & $\approx -0.22$ & P \\
6 & $\approx 15.71$ & $\approx 0.00$  & F & $\approx 15.54$ & $\approx -0.01$ & P & $\approx 15.37$ & $\approx -0.02$ & P \\
\bottomrule
\end{tabular}
\end{table}

For $v/u=0.200$, the sequence of characters (P, D, P, D, P, F) as shown in Table~\ref{tab:parity_slopes} largely maintains an alternating parity for the first five branches. The sixth branch has a slope approximately zero ($\approx 0.00$ at $\Phi/\Phi_0 = 0$), which is classified as flat/turning. While the calculated slope for branch 6 at $v/u=0.200$ is numerically $\approx 0.00$, we are not aware of an underlying symmetry in the model that would enforce an analytically exact zero slope at this specific $v/u$ ratio; its proximity to zero, however, suggests the system at $v/u=0.200$ is close to a transition point regarding the parity of the 6th mode, or that this mode is effectively non-responsive to small flux changes at $\Phi=0$. 

Interestingly, for the intermediate ratio $v/u = 0.205$, our analysis (Table~\ref{tab:parity_slopes}) indicates that the fifth and sixth branches already both exhibit paramagnetic character (slopes $\approx -0.25$ and $\approx -0.01$ respectively, for $ku \approx 12.57$ and $ku \approx 15.54$). This finding is crucial as it demonstrates that the breakdown of the simple alternating parity effect, specifically the emergence of two consecutive paramagnetic states for these higher modes, occurs for $v/u$ values at or below $0.205$.

This breakdown is further confirmed for $v/u = 0.210$, where Branches 5 and 6 are both distinctly paramagnetic (slopes $\approx -0.22$ and $\approx -0.02$, respectively). This non-alternating behavior directly supports the theoretical predictions by Deo (2021) \cite{Deo2021} concerning the failure of the simple parity rule for specific $v/u$ ratios. While our calculation identifies both anomalous states as paramagnetic (whereas Deo's illustrative context suggested both might be diamagnetic for such a breakdown \cite{Deo2021}), the crucial failure of alternation is unambiguously reproduced and now shown to initiate by $v/u = 0.205$. The slight difference in the exact magnetic character (all paramagnetic in our findings versus Deo's context) could arise from subtle differences in model parameters, definitions, or the numerical precision in identifying the exact $ku$ values at the onset of parity breakdown. However, the core physical result---the failure of simple alternation where two consecutive states exhibit the same magnetic character---remains robust.

\subsection{Persistent Current Characteristics}
This confirmed breakdown of the parity effect has potential experimental implications for persistent currents. The persistent current $I_n$ for a given branch $n$ at a specific magnetic flux $\Phi$ is proportional to the negative derivative of its energy $E_n$ with respect to flux: $I_n(\Phi) \propto -\partial E_n / \partial \Phi$. Since $E_n \propto (ku_n)^2$, it follows that $I_n(\Phi/\Phi_0) \propto -ku_n(\Phi/\Phi_0) \cdot [d(ku_n)/d(\Phi/\Phi_0)]$. The total persistent current is the sum over the contributing branches, $I_{total}(\Phi/\Phi_0) = \sum_{n=1}^{6} I_n(\Phi/\Phi_0)$.

\begin{figure}[htbp!]
\centering
\includegraphics[width=0.8\textwidth]{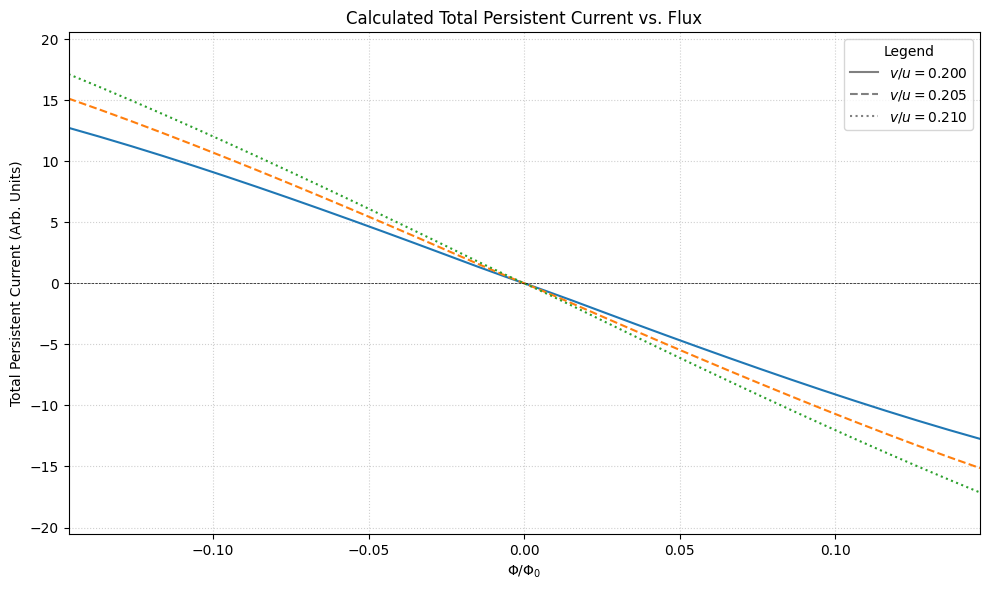} 
\caption{Calculated total persistent current (proportional to $I(\Phi) \propto -\sum_{n=1}^{6} ku_n [d(ku_n)/d(\Phi/\Phi_0)]$) versus normalized flux $\Phi/\Phi_0$ for $v/u = 0.200, 0.205,$ and $0.210$. The plot illustrates the differing behavior of the net current near zero flux due to the variations in the parity effect across the lowest six branches.}
\label{fig:persistent_current}
\end{figure}

To visually illustrate these implications, we calculated a quantity proportional to the total persistent current using the six lowest calculated dispersion branches. Figure~\ref{fig:persistent_current} shows $I_{total}(\Phi/\Phi_0)$ for the three $v/u$ ratios in the flux window $\Phi/\Phi_0 \in [-0.15, 0.15]$. 
A distinct difference in the behavior of $I_{total}$ is observed. All three $v/u$ ratios exhibit an overall diamagnetic response near zero flux, indicated by the negative slope of $I_{total}$ versus $\Phi/\Phi_0$. However, the magnitude of this response and the current values at $\Phi/\Phi_0=0$ differ. For $v/u=0.200$, $I_{total}(0) \approx 0.451$ (arbitrary units). For $v/u=0.205$ and $v/u=0.210$, where branches 5 and 6 are both paramagnetic (Table~\ref{tab:parity_slopes}), their additive positive contributions to $I_{total}(0)$ (since $s_n < 0$ for paramagnetic branches yields $I_n(0) \propto -ku_n \cdot s_n > 0$) result in higher net current values at zero flux, with $I_{total}(0) \approx 0.526$ and $\approx 0.585$ respectively. This increase in $I_{total}(0)$ for $v/u=0.205$ and $0.210$ is a direct consequence of both the 5th and 6th branches becoming paramagnetic and thus providing additive positive contributions at $\Phi/\Phi_0=0$, whereas for $v/u=0.200$ the 6th branch is flat and contributes negligibly at zero flux. The visibly steeper negative slope of $I_{total}(\Phi/\Phi_0)$ for $v/u=0.210$ compared to $v/u=0.200$ in Fig.~\ref{fig:persistent_current} directly demonstrates how the subtle change in stub length, leading to the parity effect breakdown, can substantially alter the net persistent current characteristics of the Aharonov-Bohm ring. The $v/u=0.205$ case presents an intermediate overall current response. These differences could manifest as modifications to the amplitude or even the periodicity of experimentally measured $I(\Phi)$ if higher harmonics become more dominant due to the specific sequence of state characters.

Further investigation into the behavior for $v/u$ values continuously varied around $0.200$ and $0.210$, as suggested by the reviewer, could pinpoint the exact transition to parity breakdown more precisely, though this finer scan is left for future work.

\section{Conclusion}
In this study, we have numerically solved the transcendental mode condition equation proposed by Deo (2021) \cite{Deo2021} for an Aharonov-Bohm ring with a side-attached stub, focusing on stub length to ring circumference ratios of $v/u=0.200, 0.205,$ and $0.210$. Our calculations of the dispersion relations ($ku$ vs. $\Phi/\Phi_0$), the underlying $\text{Re}(1/T)$ function, and the resulting persistent currents confirm that the system's electronic structure and magnetic response are highly sensitive to this $v/u$ ratio.

A key finding is the direct numerical evidence for a breakdown of the simple alternating parity effect. The inclusion of calculations for $v/u=0.205$ reveals that for the 5th and 6th calculated energy branches, both already exhibit paramagnetic character at this intermediate ratio. This effect persists and is clearly observed at $v/u=0.210$. Specifically, our slope analysis provides quantitative support for the theoretical discussions by Deo (2021) \cite{Deo2021} regarding parity effect breakdown, even if the precise magnetic character (paramagnetic vs. diamagnetic) of the anomalous pair differs from qualitative examples. Understanding and controlling such parity effects is crucial, as the magnetic character of energy states directly influences measurable quantities like persistent currents, as further illustrated by our direct calculation of the flux-dependent total current which varies significantly with these small changes in $v/u$. These findings could be exploited in the design of mesoscopic quantum interference devices with tailored magnetic responses. The investigation across these closely spaced $v/u$ ratios, including the intermediate $0.205$ value, helps to map the onset of this parity breakdown and underscores the intricate relationship between geometry and quantum interference in mesoscopic systems.

Future work could extend this investigation in several directions:
\begin{itemize}
    \itemsep0em 
    \item Further refine the calculation of persistent current $I(\Phi)$ across a wider flux range and for each branch $n$ to quantitatively assess the impact of the parity breakdown on this experimentally accessible observable, particularly how its harmonic content varies across the $v/u=0.200$ to $0.210$ range.
    \item Perform a finer scan of the $v/u$ parameter space, for instance, between $v/u = 0.200$ and $0.205$, to more precisely map the "phase boundary" where the parity breakdown for specific higher modes occurs.
    \item Investigate the robustness of these findings against small amounts of disorder or variations in the stub-ring coupling strength.
    \item Employ alternative numerical methods, such as transfer-matrix approaches, to cross-validate the solutions obtained via root-finding for the mode condition equation.
\end{itemize}
These computational results validate the predictive power of the underlying theoretical model and highlight the delicate tuning required to control quantum phenomena in such nanostructures.

\end{document}